\documentclass{aip-cp}

\usepackage[numbers]{natbib}
\usepackage{rotating}
\usepackage{graphicx}
\usepackage{bm}

\begin{document}

\title{Performance of CVD Diamond Single Crystals as Side-bounce Monochromators in the Laue Geometry at High Photon Energies 
\footnote{submitted for publication to AIP Conf. Proc. (Proc. of International Conference on Synchrotron Radiation Instrumentation 2018)}}

\author[aff1]{S. Stoupin \corref{cor1}}
\author[aff1]{T. Krawczyk}
\author[aff1]{J.P.C. Ruff}
\author[aff1]{K.D. Finkelstein}
\author[aff2]{H.H. Lee}
\author[aff1]{R. Huang}

\affil[aff1]{Cornell High Energy Synchrotron Source, Cornell University, Ithaca, NY 14853, USA}
\affil[aff2]{Pohang Accelerator Laboratory, Pohang 37673, Korea}
\corresp[cor1]{Corresponding author: sstoupin@cornell.edu}

\maketitle

\begin{abstract}
We report on performance of chemical vapor deposited (CVD) single crystal diamond plates as side bounce monochromators for high photon energies ($\gtrsim$~20 keV) in the Laue geometry. Several crystals were tested in-operando high-heat-load conditions at A1 undulator station of Cornell High Energy Synchrotron Source. Up to 10$\times$ enhancement in the reflected x-ray flux was observed compared to that delivered by IIa diamond plates grown by high-pressure high-temperature method. Wavefront distortions were measured using analyzer-based x-ray diffraction imaging. Focusing of a portion of the reflected beam was demonstrated using Pt-coated mono capillary optics at a photon energy of 46 keV.
\end{abstract}

\section{INTRODUCTION}

Diamond is a very attractive material for high-heat-load x-ray monochromators due to its superior thermal properties and high radiation hardness.
For a perfect crystal the fraction of radiation selected, and the resulting reflected flux are rather small due to the narrowness of the 
intrinsic angular-wavelength reflection region defined by the Darwin width. Distortions in the crystal lattice of a chemically vapor deposited (CVD) 
diamond may enable selection of a greater fraction of incoming radiation from a polychromatic and/or divergent incident radiation compared to 
that of perfect single crystals. 
Reflectivity of CVD diamond has been studied earlier for applications in neutron monochromators 
\cite{Freund09,Freund11,Fischer13}. It was found that although reflectivity for neutrons can be predictably quantified the result may 
depend on variations in the diamond microstructure, which was deduced from x-ray reflectivity measurements. 

For high-heat-load x-ray monochromators operating at high photon energies ($E \gtrsim$~20~keV) the transmission (Laue) diffraction 
geometry\footnote{In the Laue geometry the exit surface of the crystal for the reflected beam is opposite to that of the entrance surface} 
(as opposed to the reflection (Bragg) geometry) offers an advantage of limiting the incident beam footprint on the entrance crystal surface for 
the most efficient low-index reflections with shallow Bragg angles. Absolute reflectivity of CVD single crystal diamond plates was recently studied both 
theoretically (Darwin-Hamilton intensity transport equations) and experimentally (using double-crystal diffractometry and rocking curve imaging) \cite{Stoupin18}. 
It was shown that integrated reflectivity in the Laue geometry of CVD diamond crystals with a homogeneous distribution of mosaic blocks can exceed that 
of a perfect crystal by at least one order of magnitude. 
The homogeneous distribution of defects can be defined as an ensemble of mosaic blocks of certain characteristic thickness and angular misorientation. 
Availability of crystal volumes, where this description is valid was demonstrated.

In this work we study CVD diamond crystal plates as side-bounce Laue high-heat-load monochromators. Using geometric optics and basic x-ray diffraction considerations 
we derive criteria on the optimal use of imperfect crystals for this application. Several CVD plates were subjected to in-operando 
tests using synchrotron undulator radiaiton. An increase in the reflected x-ray flux of up to about one order of magnitude compared to the levels 
provided by HPHT crystal plates\footnote{IIa type diamond crystals grown using high-pressure high-temperature method} of much higher crystal quality was demonstrated. 
In-operando tests of crystal plates were performed at A1 undulator station of Cornell High Energy Synchrotron Source (CHESS). Analyzer based diffraction imaging 
yielded quantitative results on the intrinsic radiation bandwidth of the reflected beams and wavefront distortions. In addition, moderate focusing of the 
CVD-plate-reflected x-ray beam was explored using Pt-coated capillary at a high photon energy of 46~keV. The size of the focused beam was found to be close to the nominal focusing parameters of the capillary while the focused x-ray flux was about 1$\times 10^{10}$~photons/s. 

\section{GENERAL REMARKS}
A horizontal side-bounce Laue diamond monochromator is a simple solution for monochromatization of x-rays at synchrotron radiation sources. 
Among advantages of this approach is stability, affordability, and efficient use of available facility space. 
These advantages come at a price of limited energy tunability and reduced integrated throughput due to Bragg diffraction of $\pi$-polarized radiation 
(polarization of synchrotron beam in the horizontal scattering plane). 
The spectral throughput is further limited by the choice of Laue geometry. Reflectivity of a perfect crystal in the Laue geometry is an oscillating function, 
which is extremely sensitive to micron level variations in crystal thickness (e.g., \cite{Stephenson97}). In the limit of a thick nonabsorbing crystal the 
maximum reflectivity in the Laue case approaches 1/2 (e.g., \cite{Zach}). This limit is a good approximation for low order reflections for hard x-rays in diamond 
crystals of moderate thickness 0.5-1~mm.\footnote{Much thinner crystal plates as side-bounce monochromators are expected to be susceptible to heat-load-induced 
thermal distortion (due to a substantial reduction in stiffness).}
These considerations suggest that a side-bounce diamond monochromator is not the best choice for spectroscopy applications. However, for other applications 
developments of side-bounce monochromators can be aimed at optimization of the integrated throughput. A side-bounce crystal monochromator can select a greater 
fraction of photon energy bandwidth from the incident polychromatic beam, which could result in an overall increase in the reflected x-ray flux (despite reduced spectral throughput). 

\section{BANDWIDTH OF REFLECTED RADIATION. SOURCE SIZE LIMIT.}
\begin{figure}[h]
  \centerline{\includegraphics[width=350pt]{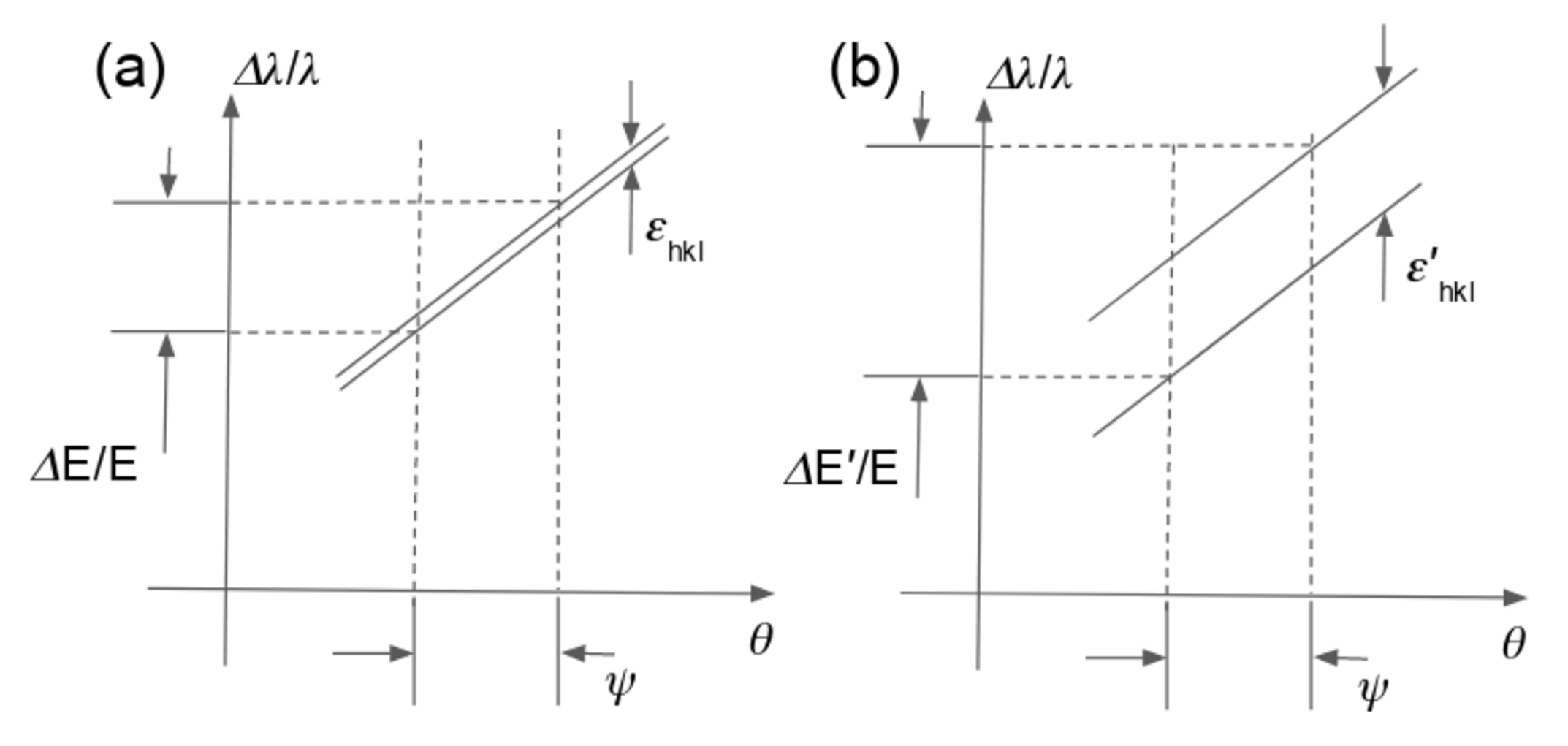}}
  \caption{Angular-wavelength reflection regions of an $hkl$ reflection: for a perfect crystal (a), and, for an imperfect crystal (b).
   The latter can be represented as a superposition of reflection regions for an ensemble of perfect crystals shifted along the angular axis ($\theta$) 
   by the amount of their angular misorientations, which are bound to a certain limiting value $\rho$.}
\label{fig:Dumd}
\end{figure}

Within the framework of geometric optics if a polychromatic x-ray beam emanating from a point source with angular divergence $\psi$ is incident on a crystal 
aligned to $hkl$ reflection, the photon energy bandwidth of the reflected x-ray beam can be approximated with

\begin{equation}
\frac{\Delta E}{E} = \sqrt{\frac{\psi^2}{\tan^2{\theta_{hkl}}} + \varepsilon^2_{hkl}},
\label{eq:dee}
\end{equation}

where $\theta_{hkl}$ is the angle corresponding to the center of the reflection region (or Bragg angle to a good approximation), 
and $\varepsilon_{hkl}$ is the intrinsic relative bandwidth of the reflection. 
If, additionally, the source has a finite size in the scattering plane $s_h$, each point on the crystal sees x-ray divergence of 

\begin{equation}
\psi_0 = \frac{s_h}{L},
\label{eq:psi0}
\end{equation}

where $L$ is the distance from source to the crystal. For estimation of the total bandwidth in Eq.~\ref{eq:dee} $\psi$ should be replaced with 
$\sqrt{\psi^2 + \psi^2_0}$. Considering horizontal source sizes of 3rd generation synchrotrons (200 - 600 $\mu$m) and source-to-monochromator 
distances of about 30 m, $\psi_0$ can be from a few to a few tens of microradians. If, additionally, high photon energies and shallow Bragg angles 
are considered $\tan{\theta_{hkl}} \simeq 0.1$, the contribution of the source size to the bandwidth can become dominant if compared to the intrinsic 
bandwidth of $hkl$ reflections in a perfect diamond crystal ($\varepsilon_{hkl} < \varepsilon_{111} \simeq 5\times 10^{-5}$).
In some cases (e.g., relatively short beamlines of CHESS $L \simeq 15$m and even greater $s_h \simeq$~1000~$\mu$m) the influence of the intrinsic 
bandwidth can be neglected. Even the reflected beam is apertured/collimated to minimize the source divergence $\psi$, the total radiation bandwidth 
is still quite considerable $\Delta E/E \simeq 1\times10^{-3}$. 
However, the smallness of $\varepsilon_{hkl}$ limits the total radiation flux selected by the reflection. The size of the reflection region in the 
angular-wavelength space for entrance waves is quite narrow as shown in Fig.~\ref{fig:Dumd}(a) (i.e., Dumond diagram). A reflection region of an imperfect 
crystal if approximated as an ensemble of misoriented blocks of effective angular misorientation $\rho \gg \varepsilon_{hkl}\tan{\theta_{hkl}}$ can be 
illustrated in the angular-wavelength space for entrance waves as a superposition of narrow reflection regions of individual blocks (see \cite{Stoupin15} for more details). 
These regions are shifted in the angular coordinate with respect to the central reflection angle, which results in a greater combined effective bandwidth 
$\varepsilon'_{hkl} \simeq \rho/\tan{\theta_{hkl}}$ (Fig.~\ref{fig:Dumd}(b)). Performance of such a crystal as a monochromator will be optimized if 

\begin{equation}
\rho \simeq \psi_0 \simeq \psi,
\label{eq:crit1}
\end{equation}
or, alternatively,
\begin{equation}
\varepsilon'_{hkl} \simeq \frac{\psi_0}{\tan{\theta_{hkl}}} \simeq \frac{\psi}{\tan{\theta_{hkl}}},
\label{eq:crit2}
\end{equation}

\section{HIGH-HEAT-LOAD SIDE-BOUNCE DIAMOND MONOCHROMATOR.}
\begin{figure}[h]
  \centerline{\includegraphics[width=300pt]{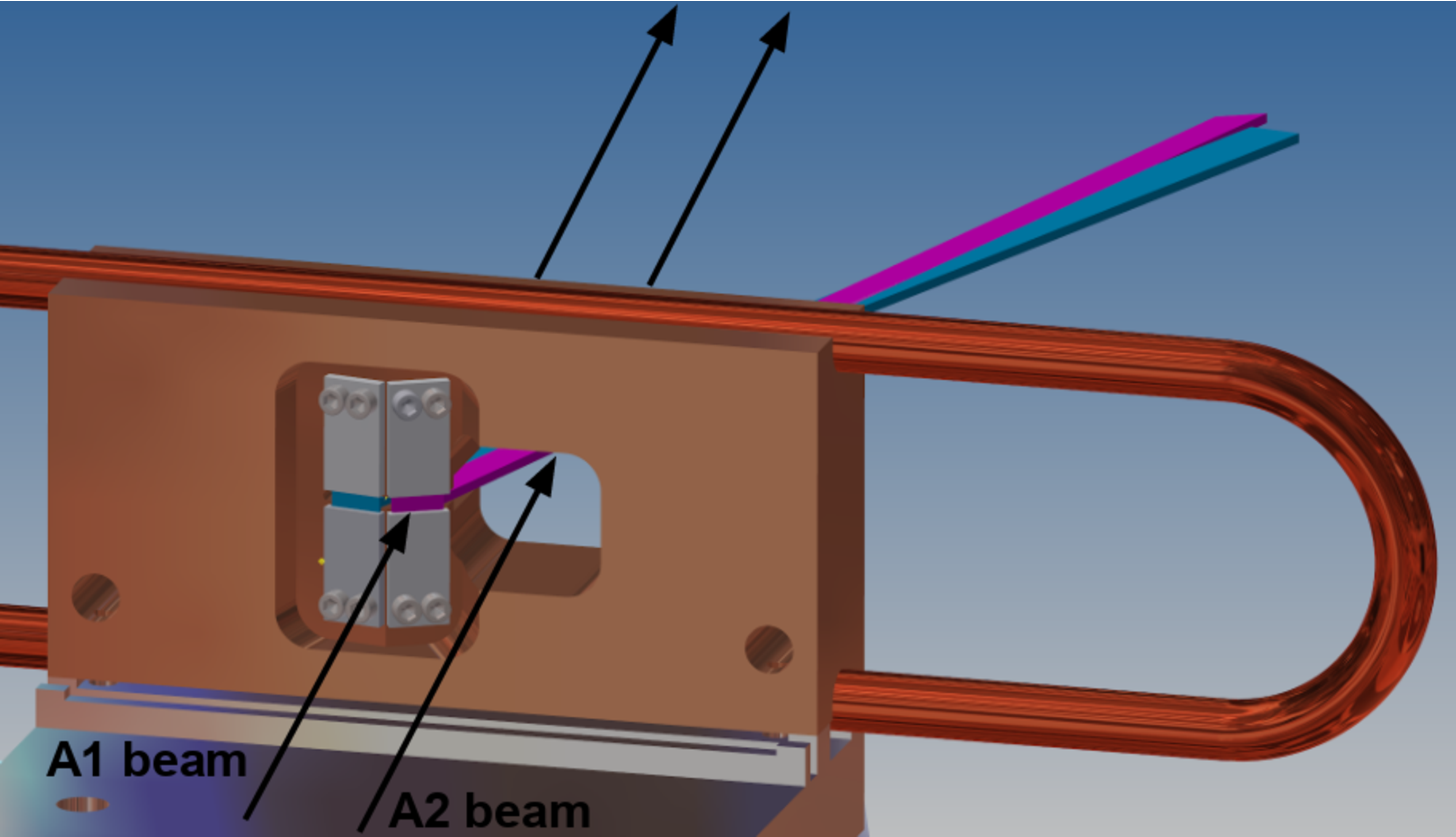}}
  \caption{Schematic of the diamond monochromator (see text for details).}
\label{fig:cmono}
\end{figure}

A pair of diamond crystal plates was installed in a water-cooled copper block of a side-bounce monochromator using the thermal mounting approach \cite{Yabashi07}. 
Aluminum fixtures were used to restrain displacements of the plates and In foil was applied on edges. 
The crystals were attached by and the thermal contact was made solely by the In foil following the procedure developed by Yabashi et al. \cite{Yabashi07}. 
The monochromator was operated at A1 undulator beamline of CHESS. The schematic of the monochromator is shown in Fig.~\ref{fig:cmono}.
At right, the undulator beam of another station (A2) passes through an opening in the water-cooled copper block while the A1 undulator beam strikes 
predominantly one of the two diamond crystals centered on the incident beam using a horizontal translation stage. 
Some overlap is unavoidable due to the tails of radiation distribution in the horizontal plane. 
The angular positions of the crystals were pre-adjusted to be in close proximity of the 111 reflection (for one of the diamond plates) 
and of the 220 reflection for the other. A small angular offset ($\simeq$~1~deg.) is introduced between the two to facilitate rejection of the unwanted 
reflection using downstream apertures. The angular range of rotation of the copper block permits final crystal alignment to the nominal scattering angle of the beamline 
$2\theta_n \simeq$17.4~deg yielding nominal working photon energies of 19.9~keV (111~reflection) and 32.5~keV (220~reflection). 
First, crystal plates CVD-1 and CVD-2 were tested in-operando conditions of the monochromator using full undulator beam delivered by CHESS compact undulator. 
In the second experiment these were replaced with crystals grown by high-pressure high-temperature method (HPHT-1 and HPHT-2) using the same 111 and 220 working reflections. These crystals, although still imperfect, had substantially smaller concentration of defects, which was verified using white beam topography (an experiment conducted at 1-BM beamline of the Advanced Photon Source). In the third experiment, the second crystal was replaced with a CVD plate set in 400 Laue reflection (CVD-3). All crystal plates had
thickness of about 0.6~mm and (001) surface orientation. The 220 and 400 were symmetric Laue reflections while the 111 reflection was slightly asymmetric 
(asymmetry angle $\simeq$~54.7~deg.) and set in the beam compressing geometry. 
The monochromator was operated at A1 station of CHESS over the course of about two years delivering monochromatic radiation for user experiments involving 
experimental techniques such as thin-film and single crystal diffraction, powder diffraction, wide-angle x-ray scattering as well as x-ray fluorescence 
mapping experiments with moderate spatial resolution ($\simeq$~25$\times$25~$\mu m^2$).

For each crystal plate/reflection, measurements of total (unapertured) flux of the reflected beams were performed using Ar filled ionization chamber. 
The results of these measurements along with theoretical estimates are shown in Table~\ref{tab:param}.
The theoretical estimates were obtained using SPECTRA software \cite{Tanaka01} to generate partial undulator flux of the CHESS compact undulator \cite{Temnykh13} and x-ray dynamical diffraction for perfect crystals combined with ray-tracing approach (assuming Gaussian shape of the x-ray beam) to calculate throughput of the monochromator. The measured values of the total flux for CVD crystal plates were consistently greater than the theoretical values for perfect crystals. These theoretical values are prone to overestimation (e.g., due to non-ideal performance of the x-ray source). 
Comparing experimentally measured flux values of for CVD and HPHT crystals a gain of about 10$\times$ is observed for the 111 reflection and about 5$\times$ for the 220 reflection. At the same time, it is expected that use of imperfect crystals as x-ray reflectors inevitably leads to distortion of radiation wavefront due to local deviations from the x-ray diffraction condition. Therefore, only a portion of the total flux can be efficiently delivered to the sample in an experiment using downstream focusing optics with a limited working aperture. Focusing aberrations due to the distorted wavefront are expected. In the following we describe studies of the radiation wavefront and results of focusing using mono capillary optics.

\begin{table}[h]
\caption{Diamond plate performance characteristics.}
\label{tab:param}
\tabcolsep7pt\begin{tabular}{lccccccc}
\hline
\tch{1}{c}{b}{Diamond \\ plate}  & \tch{1}{c}{b}{Reflection\\ $hkl$} &\tch{1}{c}{b}{Photon\\energy $[keV]$}  & \tch{1}{c}{b}{$F_t$ \tabnoteref{t1n1} \\$[ph/s]$} 
& \tch{1}{c}{b}{$F_e$\tabnoteref{t1n2}\\$[ph/s]$} &  \tch{1}{c}{b}{$\rho_m$ \tabnoteref{t1n3} \\ $[\mu rad]$}  
& \tch{1}{c}{b}{$\varepsilon'_{hkl}$\\ \ } & \tch{1}{c}{b}{$\varepsilon_{hkl}$ \\ \ } \\
\hline
CVD-1  & 111 & 19.9 & 6.5$\times$10$^{12}$  & 3.2$\times$10$^{13}$ & 71 & 4.6$\times$10$^{-4}$ & 5.4$\times$10$^{-5}$ \\
CVD-2  & 220 & 32.5 & 1.0$\times$10$^{12}$  & 2.7$\times$10$^{12}$ & 94 & 6.2$\times$10$^{-4}$ & 1.8$\times$10$^{-5}$ \\
HPHT-1 & 111 & 19.9 & 6.5$\times$10$^{12}$  & 2.3$\times$10$^{12}$ & 26 & 1.7$\times$10$^{-4}$ & 5.4$\times$10$^{-5}$ \\
HPHT-2 & 220 & 32.5 & 1.0$\times$10$^{12}$  & 5.6$\times$10$^{11}$ & 17 & 1.1$\times$10$^{-4}$ & 1.8$\times$10$^{-5}$ \\
CVD-3  & 400 & 46.0 & 1.6$\times$10$^{11}$  & 3.8$\times$10$^{11}$ & 42 & 2.7$\times$10$^{-4}$ & 7.3$\times$10$^{-6}$ \\

\hline
\end{tabular}
\tablenote[t1n1]{Total flux using theoretical calculations.}
\tablenote[t1n2]{Experimentally measured total flux.}
\tablenote[t1n3]{Average width (FWHM) of the local reflectivity curve in the $\Delta \theta_{\sigma}$ topographs (e.g., Fig.~\ref{fig:wf}(c-d))}
\end{table}

\section{WAVEFRONT CHARACTERIZATION}
\begin{figure}[th!]
  \centerline{\includegraphics[width=450pt]{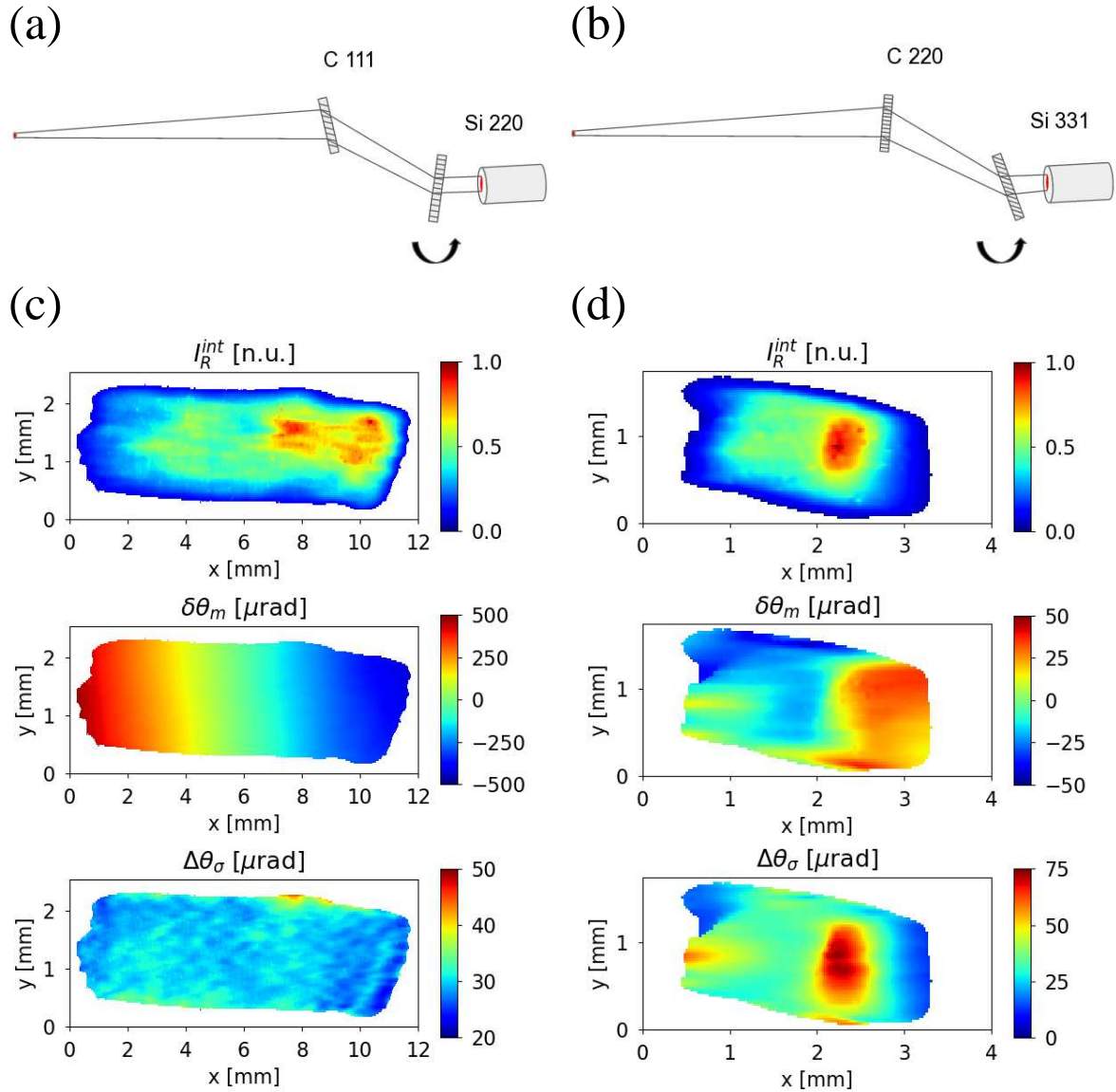}}
  \caption{Experimental arrangement for diffraction imaging of radiation wavefront in the horizontal scattering plane:
  for CVD-1 using Si 220 Laue analyzer at 19.9 keV (a), and for CVD-2 using Si 331 Laue analyzer at 32.5 keV (b).
  Analyzer rocking curve topographs: obtained using the Si 220 Laue analyzer (c), and using the Si 331 Laue analyzer (d). 
  The rocking curve topographs represent maps of the integrated reflected intensity normalized by the maximum value observed ($I^{int}_R$), 
  the curve's peak position ($\delta \theta_m$) and the curve's width as a standard deviation of a Gaussian approximation ($\Delta \theta_{\sigma}$).
  The colorbar on the $\delta \theta_m$ and $\Delta \theta_{\sigma}$ topographs are in units of $\mu$rad. }
\label{fig:wf}
\end{figure}

In-operando wavefront characterization in the horizontal scattering plane of the monochromator crystals was performed using diffraction imaging 
(\cite{Stoupin15}) in the nondispersive double-crystal arrangement, where 0.5-mm-thick Si wafer was chosen as the analyzer (second) crystal. 
For brevity, we show results only for diamond plates CVD-1 and CVD-2. The experimental arrangements for evaluation of performance of diamond crystal plates CVD-1 and CVD-2 are shown schematically in Fig.\ref{fig:wf}(a) and (b), respectively. In each case, an analyzer reflection was chosen with a d-spacing approximately matching that of the working diamond reflection. 
Sequences of images of the beam reflected from the analyzer crystal were collected using an area detector (at a distance of about 2~m from the 
monochromator) at different angular positions of the analyzer over its rocking curve. 
These images were processed on per-pixel basis to calculate rocking curve topographs representing 
maps of the integrated reflected intensity normalized by the maximum value observed ($I^{int}_R$), the curve's peak position 
($\delta \theta_m$) and the curve's width as a standard deviation of a Gaussian approximation ($\Delta \theta_{\sigma}$).
The calculated rocking curve topographs are shown in Fig.\ref{fig:wf}(c) for CVD-1 and in Fig.\ref{fig:wf}(d) for CVD-2.

The $\delta \theta_m$ topographs reveal an effect of lattice bending of the entire plate for both crystal plates. Remarkably, the horizontal gradients 
in these topographs for CVD-1 and CVD-2 have opposite signs. Radiation reflected from CVD-1 is defocused while radiation reflected from CVD-2 is focused. This is confirmed by change in the horizontal beam footprint. The size of the beam reflected from CVD-1 ($\simeq$~11~mm) in the imaging plane of the detector is about two times the size expected from geometry of the experiment in case of wavefront preservation ($\simeq$~5~mm). For CVD-2 this size is reduced to about 2~mm. The integrated reflectivity is quite nonuniform in both cases ($I^{int}_R$ topographs).
Nevertheless, the changes in the beam size are clearly identified. Using ex-situ rocking curve topography it was found that the observed lattice bending 
was intrinsically present in both crystal plates at the same levels. Thus, this effect was not dominated by induced strain due to high radiation heat load 
or crystal mounting. The topographs $\Delta \theta_{\sigma}$ showing the maps of the curve's width as standard deviation of a Gaussian approximation yield
an estimate for the angular misorientation: $\rho \approx 2.355\times \Delta \theta_{\sigma}$. The values averaged across the topographs are shown in
Table~\ref{tab:param} for all studied crystal plates. For the CVD plates they compare favorably to the source-size divergence of the experiment $\psi_0 \simeq 140 \mu$rad. The resulting intrinsic relative bandwidths $\varepsilon'_{hkl}$ are also shown in the table. For all studied plates/reflections these values were found to be greater than the intrinsic energy width of the perfect crystal $\varepsilon_{hkl}$. 

\section{FOCUSING OF REFLECTED RADIATION USING CAPILLARY OPTICS}
\begin{figure}[h]
  \centerline{\includegraphics[width=500pt]{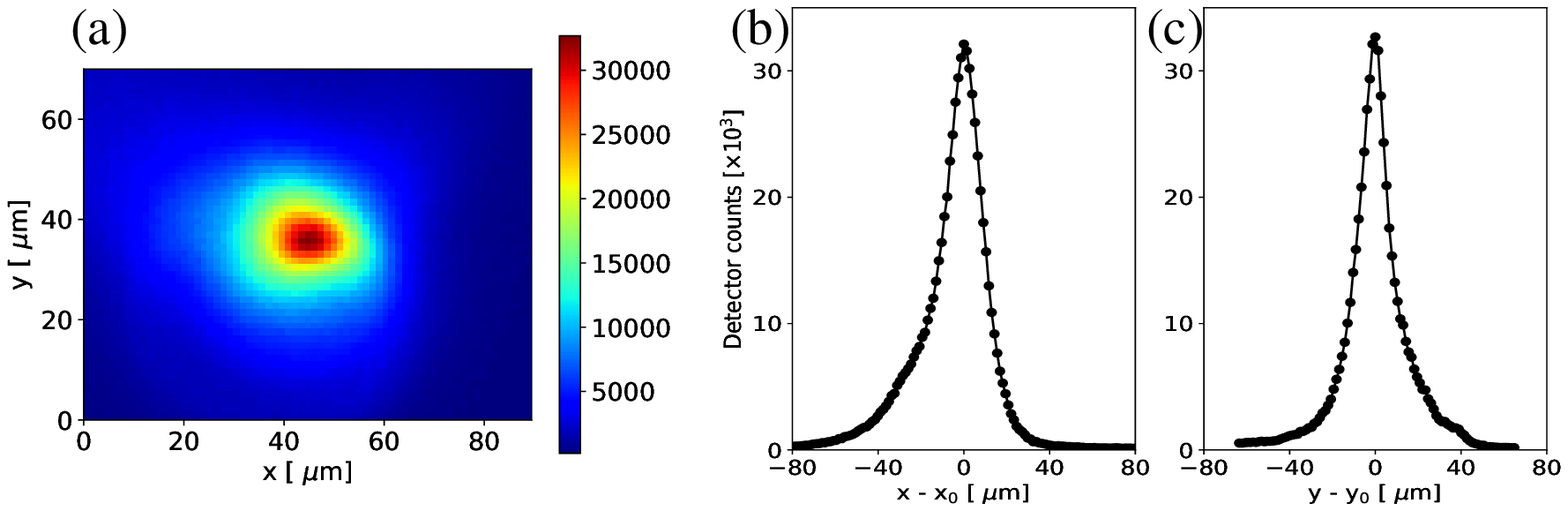}}
  \caption{Distribution of intensity in the focusing plane (a). Centered beam profiles in the horizontal (b) and vertical (c) directions (slices of the intensity distribution). The width of the horizontal profile is $\simeq$~21~$\mu$rad (FWHM) and the width of the vertical profile is $\simeq$~16~$\mu$rad (FWHM).}
\label{fig:foc}
\end{figure}

A 100-mm-long mono capillary with a nominal focal distance of 56~mm delivering a focused beam of a nominal spot size of $25 \times 25 \mu m^2$ (evaluated using optical metrology) was used for focusing x-ray beam reflected by plate CVD-3 at 46~keV. The inner surface of the capillary was coated with Pt to improve grazing-incidence reflectivity at higher photon energies. The entrance aperture of the capillary had a diameter of about 600~$\mu$m and the exit clearance aperture had a diameter of 400~$\mu$m. 
The beam entering the capillary was limited to $600 \times 600$~$\mu m^2$   using x-ray slits. The center clearance aperture was blocked using a gold bead of 400 $\mu m$ diameter centered on the optical axis of the capillary. The alignment of the capillary was optimized using an area detector placed in the focusing plane. 
The distribution of intensity captured by the detector is shown in Fig.~\ref{fig:foc}(a). Centered beam profiles in the horizontal and vertical directions (slices of the intensity distribution) are shown in Fig.~\ref{fig:foc}(b) and (c) respectively. The shapes of the curves are close to Lorentzian with an asymmetric distortion present.
A more prominent distortion is observed in the horizontal profile, which could be attributed to the wavefront distortions of the CVD diamond monochromator. However, distortions due to imperfections of the capillary can not be ruled out completely.  
The width of the horizontal profile was $\simeq$~21~$\mu$rad (FWHM) and the width of the vertical profile was $\simeq$~16~$\mu$rad (FWHM). 
The flux of the focused beam ($\simeq 1 \times 10^{10}$ photons/s) was measured using an ionization chamber placed downstream of the capillary. 
\vspace{5pt}

\section{CONCLUSIONS}
In summary, increased levels of x-ray flux reflected from CVD diamond crystal plates were demonstrated in-operando high-heat-load conditions using the plates 
as side-bounce monochromator crystals for synchrotron undulator radiation. Wavefront distortions were observed due to overall lattice bending intrinsically present in the plates. These severe wavefront distortions could be avoided using preliminary screening of diamond plates for high-heat-load monochromator application using lattice strain visualization techniques (e.g., x-ray rocking curve imaging). Moderate focusing of the reflected beams is possible. Beam spot sizes in agreement with nominal 
characteristics of a focusing element can be achieved, yet the shape of the intensity distribution in the focusing plane could be affected by the wavefront distortions.

\section{ACKNOWLEDGMENTS}
We thank our colleagues A. Lyndaker, A. Temnykh, C. Conolly, D. Pagan, A. Woll, E. Fontes and J. Brock for helpful discussions, support and encouragement. 
This work was performed in part at Cornell NanoScale Facility, an NNCI member supported by NSF Grant ECCS-1542081.
C. Alpha of CNF is acknowledged for the help and effort of capillary coating. 
Use of the Advanced Photon Source was supported by the U. S. Department of Energy, 
Office of Science, Office of Basic Energy Sciences, under Contract No. DE-AC02-06CH11357. 
This work is based upon research conducted at the Cornell High Energy Synchrotron Source (CHESS) 
which is supported by the National Science Foundation under award DMR-1332208.

\nocite{*}
\bibliographystyle{aipnum-cp}%
%

%
\end{document}